\documentclass[letterpaper, paper,11pt]{AAS}		

\usepackage{bm}
\usepackage{amsmath}
\usepackage{amssymb}
\usepackage{overcite}
\usepackage{footnpag}
\usepackage{url}
\usepackage{graphicx}
\usepackage{hyperref}
\hypersetup{
    colorlinks = true,
    linkcolor  = blue,
    citecolor  = blue,
    urlcolor   = blue
}
\usepackage{subfig}
\usepackage{caption}
\PaperNumber{26-881}

\begin{document}

\title{Beyond the cube: Overlapping Grid Methods for Debris Collision Risk Assessment}

\author{Yacob Medhin\thanks{PhD Student, Department of Aerospace Engineering, Iowa State University, IA 50011, USA. email: yacbin@iastate.edu} 
\ and Simone Servadio\thanks{Assistant Professor, Department of Aerospace Engineering, Iowa State University, IA 50011, USA. email: servadio@iastate.edu}
}

\maketitle{}

\begin{abstract}

The cube method reduces conjunction screening in orbital debris simulations to $\mathcal{O}(N)$ cost by evaluating only object pairs sharing the same grid cell at each snapshot, but systematically assigns zero collision probability to pairs separated by a cell boundary at that epoch, a failure known as boundary blindness. This paper introduces the Double Cube (DC) method, which recovers boundary-crossing conjunctions through a spatially shifted secondary grid using bin-index lookup alone, preserving $\mathcal{O}(N)$ complexity. In an isotropic convergence benchmark at $L = 50$ km, DC reduces the blindness rate from $\beta_{\mathrm{Cube}} = 7.09\%$ to $\beta_{\mathrm{DC}} = 0.55\%$, and a synchronized experiment separates the temporal component from the geometric one and leaves DC below $0.01\%$ against $0.80\%$ for the cube. Recovering these pairs changes which pairs are evaluated and leaves the probability formula untouched, so it does not improve calibration by itself, and both uncorrected methods remain substantially miscalibrated, with a reliability slope of $m = 1.1314$ for the cube and $m = 1.0996$ for DC. The formula applies the full cell volume to every evaluated pair irrespective of separation, which is the same bias characterized in the no-time-counter scheme of Direct Simulation Monte Carlo. Two corrections are derived from the geometry of the uniform cell alone: a power-law form referenced to the Robbins mean separation $\bar{d} = 0.6617L$ and a Gaussian form using both that mean and the analytically derived standard deviation $\sigma = 0.2494L$, with no quantity fitted to simulation output. The power-law correction moves the reliability slope to $0.9973$ at $k = 1$ and $0.9429$ at $k = 2$, and the Gaussian correction reaches $0.9806$ with an intercept of $0.975$, the only configuration within $3\%$ of unity on both. Both corrections have been implemented in MOCAT-MC.

\end{abstract}

\section*{Nomenclature}
 
\subsection*{Latin Symbols}
 
\begin{tabbing}
  \hspace{3.0cm} \= \kill
  $A$                          \> intercept of the log-log reliability regression; $A = 1$: perfect calibration \\[-1pt]
  $A_{ij}$                     \> binary collision flag for pair $(i,j)$ \\[-1pt]
  $a$                          \> semi-major axis of an orbit, km \\
  $c$                          \> Gaussian correction constant ($= 1.5$) \\[-1pt]
   $d_{\mathrm{gap}}$                          \>  orbital altitude band gap between two objects, km \\
  $d_{ij}$                     \> snapshot center-to-center separation of pair $(i,j)$, km \\[-1pt]
  $\bar{d}$                    \> Robbins mean separation ($= 0.6617L$), km \\[-1pt]
  $dU$                         \> cell volume denominator in collision probability, km$^{3}$ \\[-1pt]
  $dU_{\mathrm{int}}$          \> ARDC Tier~1 intersection sub-cell volume ($= L^{3}/8$), km$^{3}$ \\[-1pt]
  $dU_{\mathrm{std}}$          \> ARDC Tier~2 standard cell volume ($= L^{3}$), km$^{3}$ \\[-1pt]
  $e$                         \> orbital eccentricity \\
  $F(d_{ij};\,\mu,\sigma)$     \> Gaussian CDF of pair-distance distribution \\[-1pt]
  $f_b$                        \> empirical collision frequency in reliability bin $b$ \\[-1pt]
  $\mathbf{I}_i$               \> primary bin index of object $i$ \\[-1pt]
  $\mathbf{I}'_i$              \> shifted bin index of object $i$ \\[-1pt]
  $k$                          \> power-law correction exponent \\[-1pt]
  $L$                          \> cubic cell side length, km \\[-1pt]
  $m$                          \> log-log regression slope; $m = 1$: perfect calibration \\[-1pt]
  $N$                          \> number of objects in population \\[-1pt]
  $N_{\mathrm{obj}}$           \> number of objects per Monte Carlo seed \\[-1pt]
  $N_{\mathrm{pairs}}$         \> total evaluated pairs across all reliability bins \\[-1pt]
  $n_b$                        \> number of pairs in reliability bin $b$ \\[-1pt]
  $P_{ij}$                     \> collision probability for pair $(i,j)$ per snapshot \\[-1pt]
  $\bar{P}_b$                  \> mean predicted probability in reliability bin $b$ \\[-1pt]
  $P_{\mathrm{corr}}$          \> power-law corrected collision probability \\[-1pt]
  $P^{\mathrm{CDF}}_{\mathrm{corr}}$ \> Gaussian CDF corrected collision probability \\[-1pt]
  $P_{\mathrm{raw}}$           \> uncorrected collision probability \\[-1pt]
  $R^{2}$                      \> coefficient of determination of log-log regression \\[-1pt]
  $R_i,\,R_j$                  \> hard-body radii of objects $i$ and $j$, km \\[-1pt]
  $r_a$                          \> apogee radius $= a(1+e)$, km \\
  $r_p$                          \> perigee radius $= a(1-e)$, km \\
  $v$                            \> object speed, km/s \\
  $V_{\mathrm{rel}}$           \> relative speed of pair $(i,j)$, km/s
\end{tabbing}
 
\vspace{-0.6em}
\subsection*{Greek Symbols}
\vspace{-0.2em}
 
\begin{tabbing}
  \hspace{3.0cm} \= \kill
  $\alpha(d_{ij})$             \> per-pair correction factor, $(0,\,1]$ \\[-1pt]
  $\alpha_{\mathrm{RO}}$                          \> radial range overlap gate factor, $\{0,\,1\}$ \\
  $\beta$                      \> boundary blindness rate \\[-1pt]
  $\Delta(3)$                  \> Robbins constant for three dimensions ($= 0.661707$) \\[-1pt]
  $\Delta t_{\mathrm{cube}}$   \> grid snapshot interval, s \\[-1pt]
  $\Delta t_{\mathrm{sim}}$    \> physics sub-step interval, s \\[-1pt]
  $\Delta t_{\mathrm{trav}}$   \> mean cell traversal time ($= L/V_{\mathrm{rel}}$), s \\[-1pt]
  $\eta$                       \> normalized pair separation ($= d_{ij}/0.6617L$) \\[-1pt]
  $\mu$                        \> mean of pair-distance distribution ($= 0.6617L$), km \\[-1pt]
  $\sigma$                     \> standard deviation of pair-distance distribution ($= 0.2494L$), km
\end{tabbing}
 
\vspace{-0.6em}
\subsection*{Subscripts}
\vspace{-0.2em}
 
\begin{tabbing}
  \hspace{3.0cm} \= \kill
  $a$                          \> apogee \\
  $b$                          \> bin index in reliability diagram \\[-1pt]
  $\mathrm{corr}$              \> corrected \\[-1pt]
  $\mathrm{int}$               \> ARDC Tier~1 intersection sub-cell \\[-1pt]
  $p$                          \> perigee \\
  $\mathrm{raw}$               \> uncorrected \\[-1pt]
  $\mathrm{RO}$                          \> radial range overlap \\
  $\mathrm{std}$               \> ARDC Tier~2 standard full-cell \\[-1pt]
  $\mathrm{trav}$              \> traversal
\end{tabbing}
 
\vspace{-0.6em}
\subsection*{Abbreviations}
\vspace{-0.2em}
 
\begin{tabbing}
  \hspace{3.0cm} \= \kill
  ADR      \> Active Debris Removal \\[-1pt]
  ARDC     \> Adaptive Resolution Double Cube \\[-1pt]
  CDF      \> Cumulative Distribution Function \\[-1pt]
  Cube     \> grid-based collision screening method (Liou et al.) \\[-1pt]
  DAMAGE   \> Debris Analysis and Monitoring Architecture to the Geosynchronous Environment \\[-1pt]
  DC       \> Double Cube \\[-1pt]
  DSMC     \> Direct Simulation Monte Carlo \\[-1pt]
  ECE      \> Expected Calibration Error \\[-1pt]
  KBO      \> Kuiper Belt Object \\[-1pt]
  LEO      \> Low Earth Orbit \\[-1pt]
  LEGEND   \> LEO-to-GEO Environment Debris model \\[-1pt]
  MCS      \> Mean Collision Separation \\[-1pt]
  MOCAT-MC \> MIT Monte Carlo Orbital Capacity Assessment Tool \\[-1pt]
  NTC      \> No-Time-Counter \\[-1pt]
  PDF      \> Probability Density Function \\[-1pt]
  ROI      \> Region of Interest \\[-1pt]
  SOLEM    \> Space Objects Long-term Evolution Model
\end{tabbing}
 
\section{INTRODUCTION}
\label{sec:introduction}

The growing number of active payloads and derelict objects in Low Earth Orbit (LEO) poses a serious threat to the long-term sustainability of the near-Earth space environment~[\citen{kessler1978collision,kessler2010kessler}]. Annual assessments by the European Space Agency confirm that object densities in several altitude bands already exceed the threshold for self-sustaining collision cascades~[\citen{esa2024}]. Commercial constellation deployments have approximately doubled the active payload count in LEO over the past decade, compounding the hazard posed by existing fragmentation debris~[\citen{mcdowell2024jonathan,fcc2022spacex}]. Assessing the long-term consequences of this growth and the effectiveness of remediation strategies, including Active Debris Removal (ADR) and post-mission disposal requirements, requires evolutionary models capable of propagating large object populations over decades~[\citen{liou2011active,servadio2024risk}]. Tools of this class include the MIT Monte Carlo Orbital Capacity Assessment Tool (MOCAT-MC)~[\citen{jang2025mocat,robotics2024mit}], the NASA LEO-to-GEO Environment Debris (LEGEND) model~[\citen{liou2004legend}], and the Debris Analysis and Monitoring Architecture to the Geosynchronous Environment (DAMAGE)~[\citen{lewis2009damage,lewis2001damage}]. MOCAT-MC has been applied to dynamic risk assessment and target prioritization for Active Debris Removal missions ~[\citen{simha2025optimal}], demonstrating that accurate long-term collision probability estimates are central to identifying high-risk objects~[\citen{medhin2025jsr,medhin2025sustainability}].
 
The central computational bottleneck in these simulations is conjunction screening. Evaluating all pairs in a population of $N$ objects requires $\mathcal{O}(N^2)$ distance computations per time step, which is intractable for multi-century Monte Carlo studies involving $10^5$ or more objects~[\citen{lue2011all,george2011high}]. Liou et al.~[\citen{liou2003new}] introduced the cube method to reduce this cost, originally for asteroid-Kuiper Belt Object (KBO) conjunction assessment and later adapted to the orbital debris environment~[\citen{liou2006collision}]. The method discretizes the simulation volume into uniform cubic cells of side length $L$ and assigns a nonzero collision probability only to pairs sharing the same cell at each snapshot. Under the kinetic theory of gases assumption that objects are uniformly distributed within a cell, the collision probability for pair $(i, j)$ at snapshot $t$ is
\begin{equation}
P_{ij} = \frac{\pi (R_i + R_j)^2 \, V_{\mathrm{rel}} \, \Delta t}{dU},
\label{eq:cube_prob}
\end{equation}
where $R_i$ and $R_j$ are the object radii, $V_{\mathrm{rel}}$ is the relative speed, $dU = L^3$ is the cell volume, and $\Delta t$ is the snapshot interval. Only co-occupant pairs are evaluated, reducing conjunction screening to $\mathcal{O}(N)$~[\citen{liou2006collision}].
 
Despite this efficiency, the cube method has well-documented failure modes. Lewis et al.~[\citen{lewis2019limitations}] showed that collision rate estimates are sensitive to cell size and that a cube can produce false conjunctions between objects on non-intersecting orbital planes when their altitude separation is comparable to $L$. Diserens et al.~[\citen{diserens2020assessing}] compared cube against the Orbit Trace method across several real conjunction scenarios and found poor convergence of pairwise probability estimates and systematic over-prediction of intra-constellation rates, attributing the latter to cube's random mean anomaly sampling. The most fundamental limitation is spatial \emph{boundary blindness}: when two objects are proximate but fall in adjacent cells at the snapshot epoch, Equation~(\ref{eq:cube_prob}) is never evaluated, and the collision probability is recorded as exactly zero.
 
\begin{figure}[ht]
  \centering
  \includegraphics[width=\linewidth]{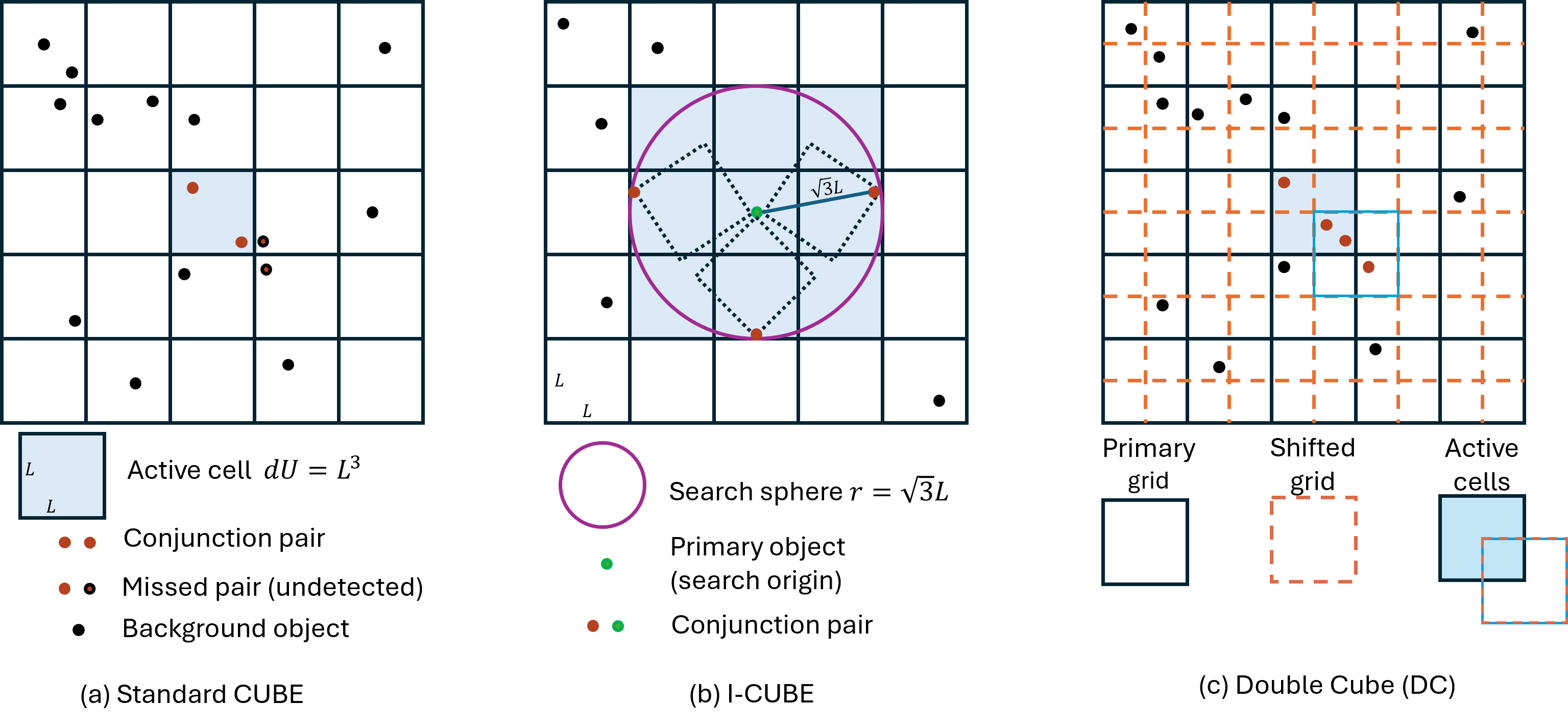}
  \caption{Comparison of three grid-based collision screening approaches. (a)~Standard cube evaluates pairs sharing the same cell ($dU = L^3$); pairs in adjacent cells receive zero probability. (b)~Improved-Cube (I-Cube) searches a sphere of radius $\sqrt{3}L$, increasing the evaluated candidate count by roughly a factor of twenty-two. (c)~Double Cube (DC) uses a shifted secondary grid offset by $L/2$ to recover the same pairs through bin-index lookup alone, preserving $\mathcal{O}(N)$ complexity. Solid and dashed outlines denote the primary and shifted grids, respectively; shaded regions indicate the active cells that contain at least one object. Marker styles and colors in panel~(a) are consistent across
  all panels.}
  \label{fig:method_comparison}
\end{figure}

Several approaches recover missed conjunctions, each at a high computational cost. The Orbit Trace method evaluates the geometric intersection of orbital paths at $\mathcal{O}(N^2)$ per time step~[\citen{diserens2020assessing}]. The Smart Sieve method~[\citen{alarcon2002collision}] reduces the pair list through distance-based filters, but identifying qualifying pairs still requires Euclidean computation, reintroducing pairwise scaling in dense regimes. The Improved-Cube method, used in the Space Objects Long-term Evolution Model (SOLEM)~[\citen{wang2019introduction,wang2019icube}], replaces the grid cell with a search sphere of radius $\sqrt{3}L$ and adjusts the denominator to that sphere's volume, $4\sqrt{3}\pi L^3 \approx 21.8\, L^3$, which increases the number of evaluated candidates by roughly a factor of twenty-two and requires explicit distance computation against the surrounding population. Figure~\ref{fig:method_comparison} illustrates the detection geometry of each approach.

The cube algorithm makes two decisions for each pair at every snapshot, whether to evaluate the pair at all and what probability to assign it, and the errors at the two levels are independent. At the candidacy level, the co-occupancy rule admits pairs whose orbits cannot intersect and omits pairs that straddle a cell boundary, and no modification of Equation~(\ref{eq:cube_prob}) repairs either mode, since in the first case the formula is applied to a pair that should have been excluded and in the second it is never applied at all. At the value level, Equation~(\ref{eq:cube_prob}) applies the full cell volume $L^3$ to every evaluated pair irrespective of where within the cell the two objects lie, so a pair at contact and a pair separated by nearly a full cell diagonal receive the same probability. The same expression governs molecular collision events in Direct Simulation Monte Carlo (DSMC) gas dynamics as the No-Time-Counter (NTC) algorithm~[\citen{bird1994}], and Alexander, Garcia, and Alder~[\citen{alexander1998}] showed that transport coefficients computed under this scheme are overestimated by a term scaling as the square of the normalized cell size. For two points drawn uniformly from a cube of side $L$, the expected center-to-center separation is $\bar{d} = 0.6617\, L$, and any detected pair whose actual separation $d_{ij}$ exceeds this value is assigned too high a probability. The DC method addresses the candidacy level, and the corrections derived in the methodology section address the value level.
 
This paper introduces the Double Cube (DC) method, which recovers boundary-crossing conjunctions through a spatially shifted secondary grid using bin-index lookup alone, preserving $\mathcal{O}(N)$ complexity. The DC method offsets a secondary grid by $L/2$ in each Cartesian dimension; any pair separated by a primary-grid boundary on a single axis shares a cell in the shifted grid along that axis, requiring only two bin-index lookups per object without calculating the Euclidean distance. The Adaptive Resolution Double Cube (ARDC) is an instrumentation variant that applies a reduced volumetric denominator to pairs confirmed by both grids simultaneously. The resulting overestimation, which grows with the degree of volume reduction, motivates the corrections derived here. The first correction is a power-law function of the ratio of the detected pair's snapshot separation to the Robbins mean distance $\bar{d} = 0.6617\, L$, evaluated at the DSMC-motivated exponents $k = 1$ and $k = 2$. The second is derived entirely from the first and second moments of the pair-distance distribution, both obtained analytically from the uniform cell geometry. Both methods are validated against a deterministic physics engine in an isotropic Rush-In convergence scenario at $L = 50$ km. DC reduces the blindness rate from $\beta_{\mathrm{Cube}} = 7.09\%$ to $\beta_{\mathrm{DC}} = 0.55\%$, and a synchronized experiment isolates the geometric residual to $0.80\%$ for the cube against less than $0.01\%$ for DC. The power-law correction moves the reliability slope from $1.0996$ to $0.9973$ at $k = 1$ and $0.9429$ at $k = 2$, and the parameter-free Gaussian correction reaches $0.9806$ with an intercept of $0.975$, the only configuration within $3\%$ of unity on both metrics.
 
The paper is organized as follows. The methodology section presents the DC and ARDC architectures and derives both correction frameworks and the calibration metrics. The simulation environment section describes the Rush-In scenario, the physics engine, and the Monte Carlo protocol. The results section reports blindness, pair-distance, and calibration findings. The orbital capacity section presents the MOCAT-MC implementation, the radial range overlap gate, and representative 50-year ensemble projections. The conclusion summarizes the findings.

\section{METHODOLOGY: DISCRETIZED COLLISION RISK ASSESSMENT}
\label{sec:methodology}
 
\subsection{Baseline Spatial Discretization}
\label{subsec:cube_baseline}

The standard cube algorithm maps the continuous three-dimensional position vector of each object, $\mathbf{r}_i = [x_i,\, y_i,\, z_i]^T$, to a discrete bin index. For a grid cell of side length $L$, the primary bin index $\mathbf{I}_i$ is computed as
\begin{equation}
\mathbf{I}_i = \left\lfloor \frac{\mathbf{r}_i}{L} \right\rfloor,
\label{eq:bin_index}
\end{equation}
where $\lfloor\cdot\rfloor$ denotes the component-wise floor operation. Collision risk is evaluated only for pairs $(i,j)$ satisfying $\mathbf{I}_i = \mathbf{I}_j$. When this condition holds, $P_{ij}$ is computed from Equation~(\ref{eq:cube_prob}) with $dU = L^3$. Otherwise, the assigned probability is exactly zero. A pair of objects that are physically proximate but separated by a grid boundary at the snapshot epoch is therefore invisible to the algorithm, regardless of their true center-to-center separation, producing the boundary-blindness failure mode described in the introduction.

\subsection{The DC Architecture}
\label{subsec:dc_architecture}
 
The DC method recovers boundary-crossing conjunctions by superimposing a secondary grid over the primary one. The secondary grid is offset by $L/2$ along all three Cartesian axes, so that any pair separated by a primary-grid boundary on a single axis shares a cell in the shifted grid along that axis. The shifted bin index $\mathbf{I}'_i$ is
\begin{equation}
\mathbf{I}'_i = \left\lfloor \frac{\mathbf{r}_i - L/2}{L} \right\rfloor.
\label{eq:shifted_bin_index}
\end{equation}
A pair $(i,j)$ is flagged for probability evaluation if it shares a cell in either grid:
\begin{equation}
\left(\mathbf{I}_i = \mathbf{I}_j\right) \quad \text{or} \quad \left(\mathbf{I}'_i = \mathbf{I}'_j\right).
\label{eq:dc_detection}
\end{equation}
When Equation~(\ref{eq:dc_detection}) is satisfied, $P_{ij}$ is computed from Equation~(\ref{eq:cube_prob}) using the standard cell volume $dU = L^3$. Detection costs two bin-index lookups per object, computed from position vectors already available at each snapshot, with no Euclidean distance evaluation. Figure~\ref{fig:dc_grid} illustrates this geometry for a two-dimensional cross-section of the dual-grid system.

The $L/2$ offset places every primary-grid boundary at the center of a shifted cell, so on any single axis the two grids cannot both separate a pair closer than half a cell. This one-dimensional complementarity does not extend to three dimensions, where detection requires one grid to hold the pair on all three axes simultaneously, and the primary grid may hold two axes and fail the third exactly where the shifted grid holds the third and fails another. The residual is small and is quantified in the results section.

\subsection{Adaptive Resolution Double Cube (ARDC)}
\label{subsec:ardc}
 
\begin{figure}[ht]
    \centering
    \includegraphics[width=0.40\textwidth]{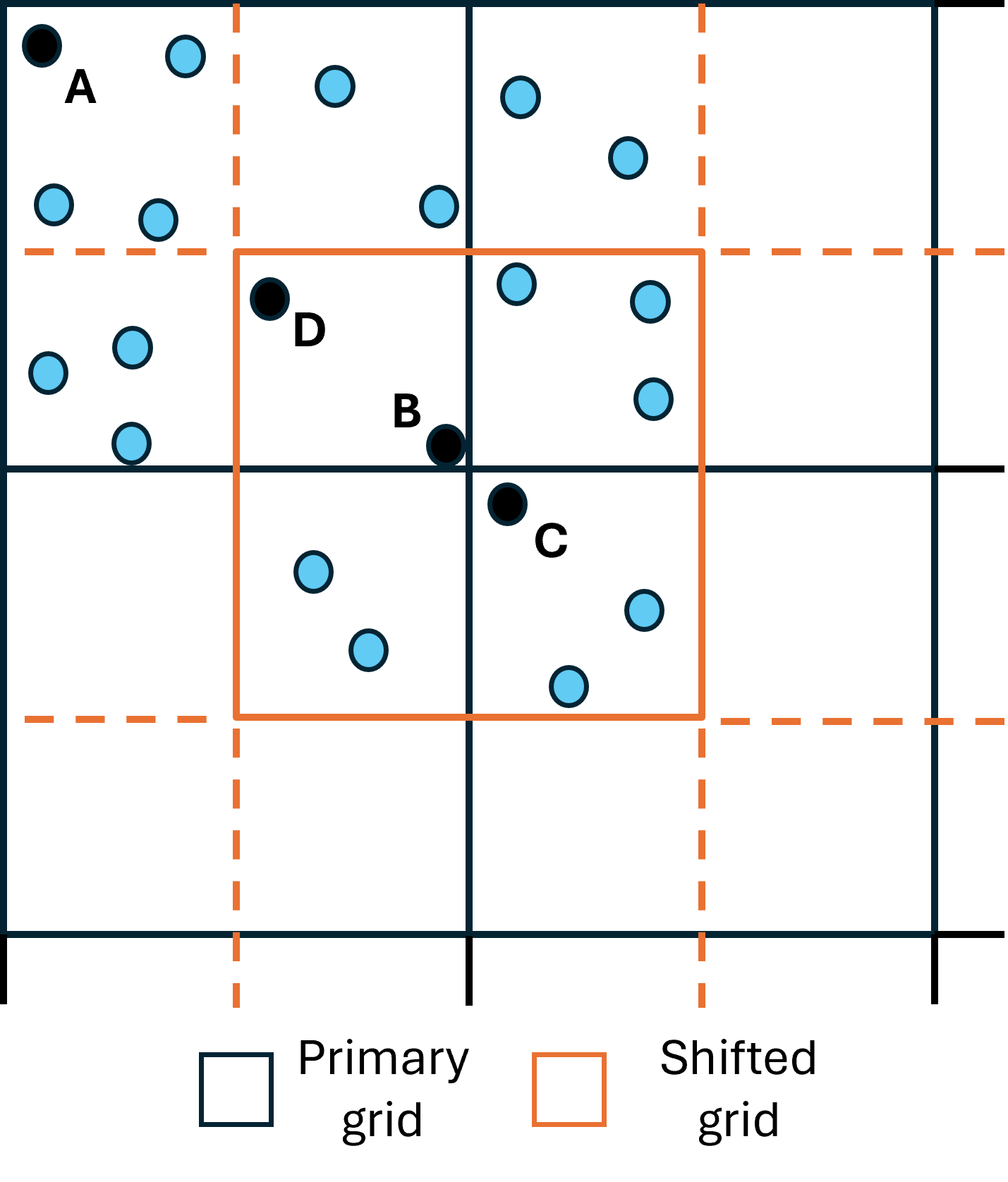}
    \caption{DC, dual-grid geometry. The primary grid (black) and the shifted secondary grid (orange) are offset by $L/2$. Objects B and C occupy adjacent primary cells and are missed by the cube, but share a cell in the shifted grid and are recovered. Object D satisfies both conditions of Eq.~(\ref{eq:dc_detection}) simultaneously and is evaluated once.}
    \label{fig:dc_grid}
\end{figure}

The ARDC variant uses the same dual-grid detection logic as DC but adjusts the volumetric denominator $dU$ in Equation~(\ref{eq:cube_prob}) based on which detection condition fired. When a pair satisfies both grid conditions simultaneously, that is $(\mathbf{I}_i = \mathbf{I}_j)$ and $(\mathbf{I}'_i = \mathbf{I}'_j)$, the two objects are geometrically confined to the intersection of their respective primary and shifted cells. For an $L/2$ offset, this intersection sub-volume has side length $L/2$, giving a volume of $(L/2)^3 = L^3/8$. The ARDC variant therefore applies a two-tier volumetric rule:
 
\begin{itemize}
\item \textbf{Tier 1 (intersection):} $(\mathbf{I}_i = \mathbf{I}_j)$ and $(\mathbf{I}'_i = \mathbf{I}'_j)$. The pair is geometrically confined to the sub-cell of side length $L/2$ formed by the overlap of the two grids, as illustrated by object D in Figure~\ref{fig:dc_grid}. Applied volume: $dU_{\mathrm{int}} = (L/2)^3 = L^3/8$.
\item \textbf{Tier 2 (single-grid):} Exactly one of the two grid conditions holds. The pair occupies adjacent primary grid cells and is recovered by one grid alone, as illustrated by objects B and C in Figure~\ref{fig:dc_grid}. Applied volume: $dU_{\mathrm{std}} = L^3$.
\end{itemize}
 
The Tier 1 volume is eight times smaller than $dU_{\mathrm{std}}$, increasing the computed $P_{ij}$ by the same factor for those pairs relative to the standard cube estimate. A sensitivity analysis with intermediate Tier 1 volumes of $L^3/2$ and $L^3/4$ showed monotonic improvement in calibration as the volume increased from $L^3/8$ to $L^3$, confirming that the overestimation worsens as the volumetric reduction becomes more aggressive. Volume reduction alone therefore does not calibrate the estimate, which motivates the first-principles corrections derived in the following two sections.

\subsection{Mean Collision Separation (MCS) Correction}
\label{subsec:mcs_correction}
 
Reducing boundary blindness through DC does not, by itself, produce a calibrated probability estimate. Equation~(\ref{eq:cube_prob}) applies the full cell volume $dU = L^3$ to every detected pair, embedding an assumption about where within the cell the two objects lie, which is the source of the systematic overestimation.
 
Equation~(\ref{eq:cube_prob}) is mathematically identical to the No-Time-Counter (NTC) collision algorithm of the standard DSMC94 method~[\citen{bird1994}], which selects collision partners randomly within each cell and computes the collision probability using the full cell volume as the denominator. Alexander, Garcia, and Alder~[\citen{alexander1998,alexander2000erratum}] analyzed the accuracy of DSMC94 and showed that the resulting transport coefficient errors scale quadratically with cell size as $(\Delta x / \lambda)^2$, establishing the theoretical basis for the overestimation mechanism. Gallis, Torczynski, Rader, and Bird~[\citen{gallis2009}] and Bird et al.~[\citen{bird2009}] subsequently identified the Mean Collision Separation (MCS) as the physical diagnostic underlying this error: in DSMC94 with random partner selection within a cell of side $L$, the expected center-to-center distance between any two collision candidates is
\begin{equation}
\bar{d} = 0.6617\,L.
\label{eq:robbins}
\end{equation}
This value is the expected distance between two points drawn uniformly at random from a cube of side $L$~[\citen{gallis2009,bird2009,gallis2011}]. Any detected pair whose actual snapshot separation $d_{ij}$ exceeds $\bar{d}$ is therefore assigned too high a collision probability, since Equation~(\ref{eq:cube_prob}) applies a cell volume calibrated for the mean separation regardless of the pair's actual distance~[\citen{gallis2009,bird2009}]. This bias is present in the cube and in DC alike, since the two differ only in which pairs enter the candidate set and share Equation~(\ref{eq:cube_prob}) unchanged. The power-law correction derived below addresses this bias, and the Gaussian correction section derives a second, independent correction from the geometry of the pair-distance distribution itself.
 
The Robbins prediction in Equation~(\ref{eq:robbins}) is validated against simulation data using an independent diagnostic before it is used as the correction denominator. For each detected pair across all Monte Carlo seeds and snapshot intervals, the Euclidean separation $d_{ij}$ is recorded at detection time, and its empirical mean is computed. This measured mean is then compared directly to $0.6617\, L$. The agreement between the measured and predicted values, including the physical explanation for any discrepancy arising from the Rush-In initialization geometry, is reported in the results section.
 
Motivated by the DSMC analysis, the normalized pair separation is defined as
\begin{equation}
\eta = \frac{d_{ij}}{0.6617\,L}
\label{eq:eta}
\end{equation}
and the per-pair correction is applied as
\begin{equation}
P_{\mathrm{corr}} = P_{\mathrm{raw}} \cdot \min\!\left(\frac{1}{\eta^k},\; 1\right).
\label{eq:mcs_correction}
\end{equation}
When $\eta > 1$, the detected pair is more separated than the Robbins mean, and Equation~(\ref{eq:cube_prob}) has assigned too high a probability; the factor $1/\eta^k$ reduces $P_{\mathrm{raw}}$ accordingly. When $\eta \leq 1$, the pair is closer than the mean, and the $\min(\cdot,\,1)$ ceiling preserves $P_{\mathrm{raw}}$ unchanged, preventing inflation of already close-approach estimates. The correction in Equation~(\ref{eq:mcs_correction}) requires only $d_{ij}$, derived from the same position vectors already used to evaluate $V_{\mathrm{rel}}$ in Equation~(\ref{eq:cube_prob}); no additional data collection or auxiliary loops are required, and the per-pair cost remains $\mathcal{O}(1)$.
 
The exponent $k$ governs the rate at which the correction attenuates with increasing pair separation. The value $k = 2$ is motivated by two complementary results from the DSMC literature. First, Alexander, Garcia, and Alder~[\citen{alexander1998,alexander2000erratum}] established that transport coefficient errors in DSMC94 NTC scale as the square of the normalized cell size, $(\Delta x / \lambda)^2$, indicating that a quadratic correction is the physically appropriate form for addressing cell-size-induced overestimation. Second, Gallis et al.~[\citen{gallis2009, gallis2008}] and Bird et al.~[\citen{bird2009}] demonstrated through their MCS analysis that reducing the collision-partner separation below the DSMC94 value of $0.6617\, L$ produces proportional reductions in transport coefficient error, confirming that the normalized separation ratio $\eta = d_{ij} / \bar{d}$ is the correct diagnostic variable and that a second-order dependence on $\eta$ is physically grounded. The case $k = 1$ provides the bounding linear correction. Together, the two values span the range of correction strengths, and both are selected on theoretical grounds alone. The results for $k = 1$ and $k = 2$ are presented and compared in the results section.

\subsection{Gaussian Pair-Distance Correction}
\label{subsec:gaussian_correction}
 
The power-law correction of the MCS correction section draws on the DSMC kinetic theory connection and uses two bounding exponents selected on theoretical grounds. A second, independent correction can be constructed from the geometry of the pair-distance distribution itself, using only the analytically derived parameters of that distribution. The derivation proceeds from the first and second moments of the distribution of center-to-center separations between two points drawn uniformly from a cube of side $L$.
 
The mean separation between two such points equals the mean distance
between two points drawn uniformly and independently from a unit cube,
scaled by $L$. This quantity is the Robbins constant for $n = 3$
dimensions~[\citen{robbins1978}], $\Delta(3) = 0.661707$. The DSMC
literature expresses it as the six-dimensional integral over the unit
cube~[\citen{gallis2008}]

\begin{equation}
\mu = L \int_0^1\!\int_0^1\!\int_0^1\!\int_0^1\!\int_0^1\!\int_0^1
  \sqrt{(x_1-y_1)^2 + (x_2-y_2)^2 + (x_3-y_3)^2}
  \;dx_1\,dx_2\,dx_3\,dy_1\,dy_2\,dy_3.
\label{eq:mu_integral}
\end{equation}

\begin{figure}[ht]
    \centering
    \subfloat[Pair-distance PDF: empirical (grey fill), analytically
        derived Gaussian with $\mu=0.6617L$, $\sigma=0.2494L$ (solid),
        and fitted Gaussian (dashed).]{%
        \includegraphics[width=0.49\textwidth]{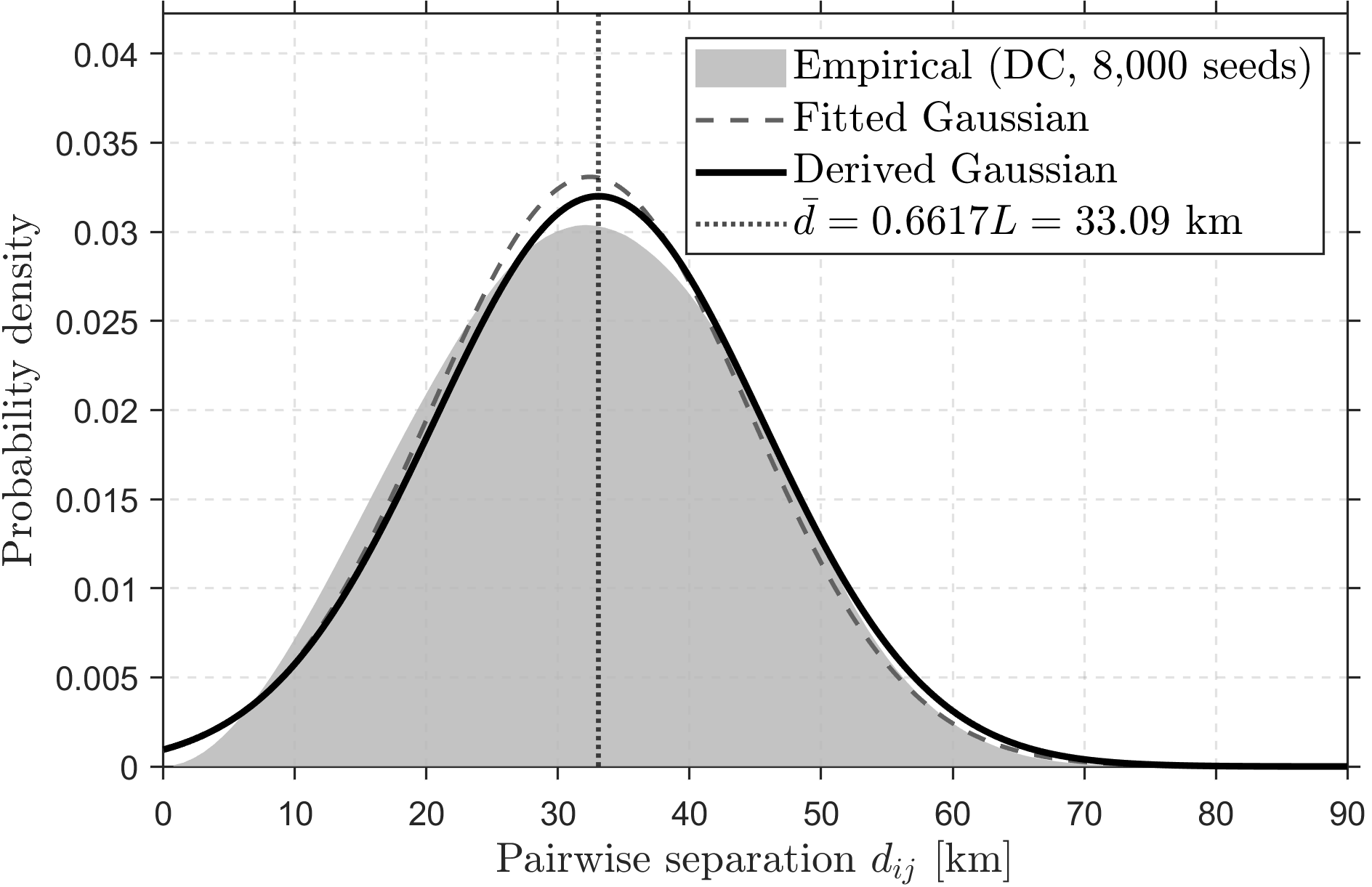}%
        \label{fig:gaussian_fit_a}%
    }%
    \hfill%
    \subfloat[Absolute CDF deviation for derived (solid) and fitted
        (dashed) parameters; dotted horizontal line marks the $0.02$
        RMSE acceptance threshold.]{%
        \includegraphics[width=0.49\textwidth]{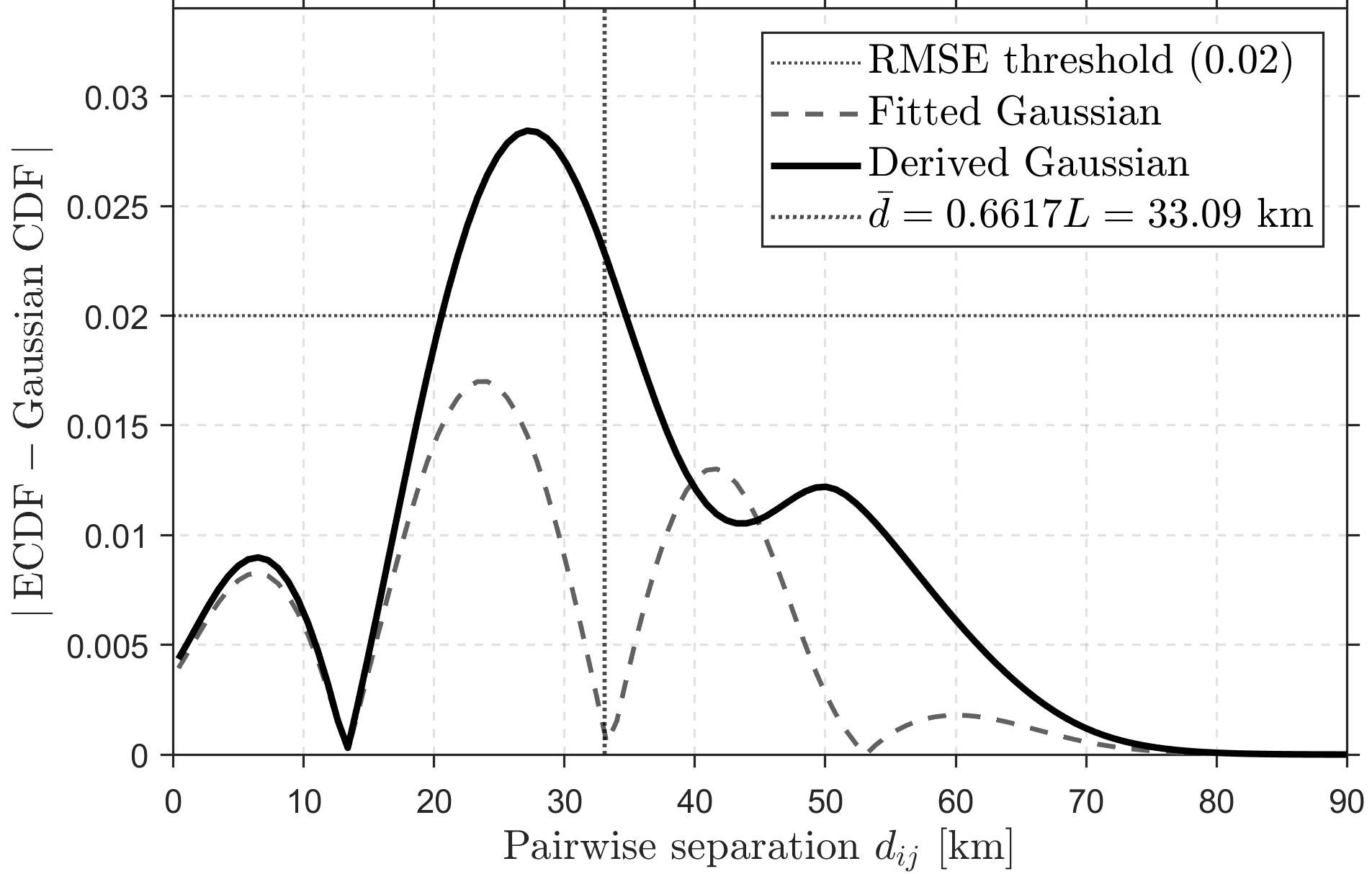}%
        \label{fig:gaussian_fit_b}%
    }%
    \caption{Pair-distance distribution for DC-detected pairs across
        $8{,}000$ Rush-In Monte Carlo seeds. The dotted vertical line
        in both panels marks the Robbins mean
        $\bar{d} = 0.6617\,L = 33.09$ km.}
    \label{fig:gaussian_fit}
\end{figure}

Equation~(\ref{eq:mu_integral}) is evaluated numerically in Table~1
of~[\citen{gallis2008}], which reports $\Delta(3) = 0.661707$
and confirms that this value is independent of the number of simulators
per cell (Table~2 of~[\citen{gallis2008}]). The corresponding
mean pair separation in a cell of side $L$ is therefore~[\citen{gallis2009,bird2009,gallis2011}]
\begin{equation}
\mu = \bar{d} = 0.6617\,L.
\label{eq:mu_robbins}
\end{equation}
This is the Robbins mean separation of Equation~(\ref{eq:robbins}), now identified as the first moment of the pair-distance distribution under the uniform cell assumption. Figure~\ref{fig:gaussian_fit_a} shows the empirical pair-distance distribution across all $8{,}000$ Monte Carlo seeds alongside both the analytically derived Gaussian and a Gaussian fitted directly to the data.
 
The second moment of the pair-distance distribution gives an exact closed-form result that requires no numerical integration~[\citen{servadio2020recursive}]. By linearity of expectation and the independence of the three coordinate differences,
\begin{align}
\mathrm{E}[d^2]
&= \mathrm{E}\!\left[(x_1-x_2)^2 + (y_1-y_2)^2 + (z_1-z_2)^2\right]
 = 3\,\mathrm{E}\!\left[(x_1-x_2)^2\right].
\label{eq:second_moment_decomp}
\end{align}
For $x_1, x_2 \sim \mathrm{Uniform}[0, L]$, the coordinate difference $\Delta x = x_1 - x_2$ has zero mean and variance
\begin{equation}
\mathrm{Var}(\Delta x)
= \mathrm{Var}(x_1) + \mathrm{Var}(x_2)
= \frac{L^2}{12} + \frac{L^2}{12}
= \frac{L^2}{6},
\label{eq:coord_variance}
\end{equation}
so $\mathrm{E}[(\Delta x)^2] = L^2/6$. Substituting into Equation~(\ref{eq:second_moment_decomp}),
\begin{equation}
\mathrm{E}[d^2] = 3 \times \frac{L^2}{6} = \frac{L^2}{2}.
\label{eq:second_moment}
\end{equation}
The standard deviation of the pair-distance distribution then follows from the variance decomposition $\mathrm{Var}(d) = \mathrm{E}[d^2] - \mu^2$:
\begin{equation}
\sigma = \sqrt{\mathrm{E}[d^2] - \mu^2}
       = \sqrt{\frac{L^2}{2} - (0.6617\,L)^2}
       = L\sqrt{\frac{1}{2} - 0.6617^2}
       = L\sqrt{0.06215}
       \approx 0.2494\,L.
\label{eq:sigma_derivation}
\end{equation}

Both $\mu$ and $\sigma$ are derived from the geometry of the uniform cell distribution alone. No simulation data, calibration runs, or free parameters enter either expression.
 
With both parameters determined analytically, the cumulative distribution function (CDF) of the pair-distance for a detected pair at separation $d_{ij}$ is approximated as Gaussian:
\begin{equation}
F(d_{ij};\,\mu,\,\sigma)
= \frac{1}{2}\!\left[1 + \mathrm{erf}\!\left(
  \frac{d_{ij} - \mu}{\sigma\sqrt{2}}\right)\right].
\label{eq:gaussian_cdf}
\end{equation}
At $d_{ij} = \mu$, Equation~(\ref{eq:gaussian_cdf}) evaluates to exactly $F(\mu) = 0.5$ by the symmetry of the Gaussian about its mean. The correction derived below is built on this property. Figure~\ref{fig:gaussian_fit_b} quantifies this agreement: the CDF root-mean-square error of the analytically derived parameters is $0.013$, against $0.007$ for the fitted Gaussian.

The correction factor $\alpha(d_{ij}) \in (0,1]$ is constructed to satisfy two requirements: it must equal unity for pairs at or below the Robbins mean (no suppression for close pairs) and decrease monotonically for $d_{ij} > \mu$. Consider the family $\alpha = \min(c - F(d_{ij};\,\mu,\,\sigma),\, 1)$, where $c$ is a constant. Applying the neutrality requirement at $d_{ij} = \mu$ gives
\begin{equation}
\alpha(\mu) = 1 \implies c - F(\mu) = 1 \implies c = 1 + F(\mu) = 1 + 0.5 = 1.5.
\label{eq:c_derivation}
\end{equation}
The constant $c = 1.5$ is therefore not a free parameter; it is determined uniquely by the requirement that the correction is neutral at the Robbins mean. Substituting into the family gives the Gaussian pair-distance correction:

\begin{equation}
P_{\mathrm{corr}}^{\mathrm{CDF}}
= P_{\mathrm{raw}} \cdot \min\!\left(1.5 - F(d_{ij};\,\mu,\,\sigma),\; 1\right).
\label{eq:cdf_correction}
\end{equation}

\begin{figure}[ht]
\centering
\includegraphics[width=0.8\textwidth]{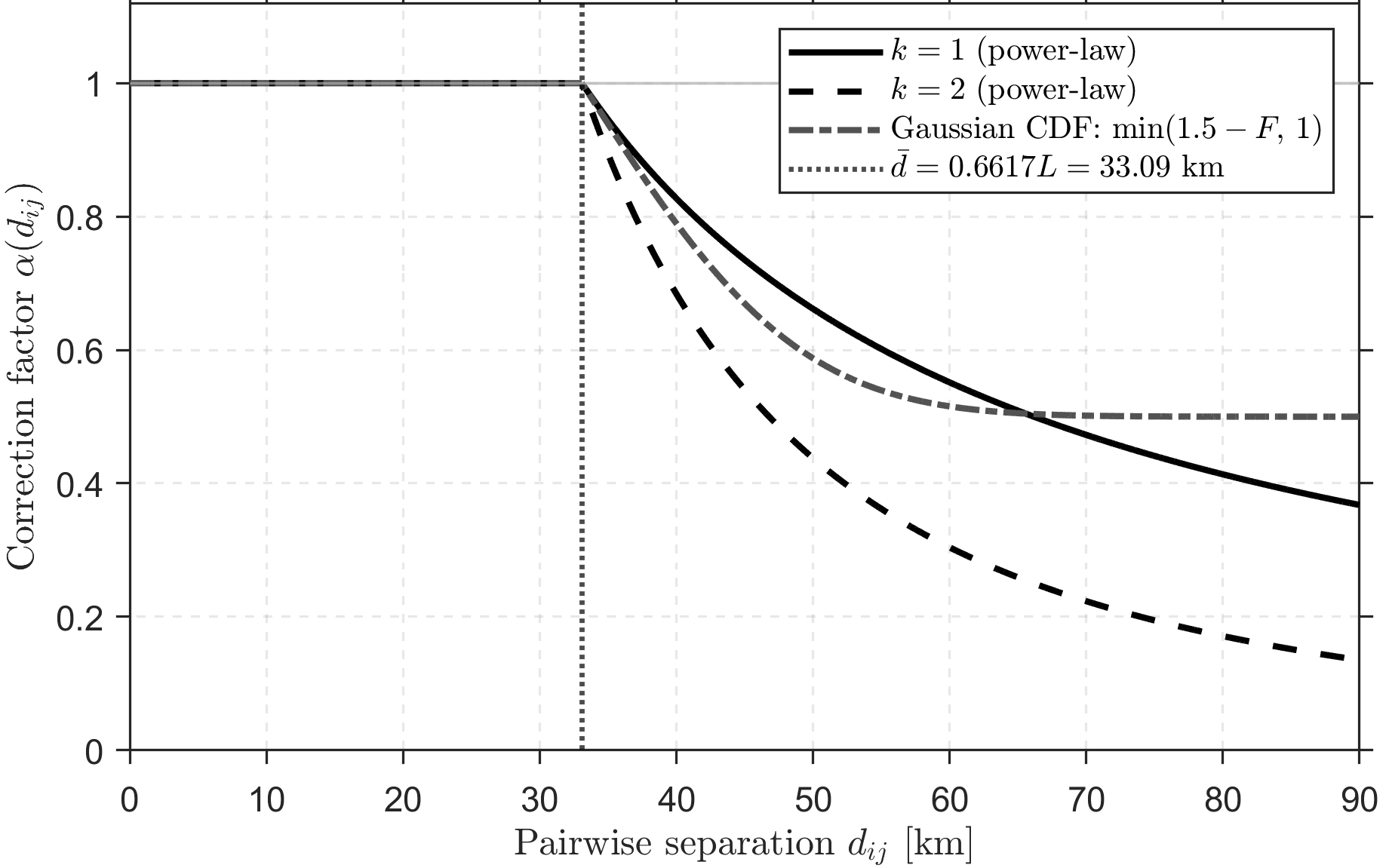}
\caption{Correction factor $\alpha(d_{ij})$ for the three variants:
power-law at $k = 1$ (solid), power-law at $k = 2$ (dashed), and
Gaussian CDF (dash-dot).}
\label{fig:alpha_profiles}
\end{figure}

For $d_{ij} < \mu$: $F < 0.5$, so $1.5 - F > 1$, but the $\min(\cdot,1)$ cap holds $\alpha = 1$, and close pairs receive no suppression and no inflation. For $d_{ij} = \mu$: $F = 0.5$ exactly, so $\alpha = 1.0$. For $d_{ij} > \mu$: $F > 0.5$, so $\alpha < 1$, and the suppression increases monotonically with distance. Figure~\ref{fig:alpha_profiles} illustrates the correction factor profiles for Equation~(\ref{eq:cdf_correction}) alongside the power-law correction of Equation~(\ref{eq:mcs_correction}) at $k = 1$ and $k = 2$, showing that the Gaussian CDF correction occupies the same correction range but with a distribution-aware shape that suppresses more gradually near $\mu$ and more strongly for pairs well beyond it.

Table~\ref{tab:gaussian_params} compares the analytically derived parameters to those measured directly from the Rush-In simulation data. The empirical mean and standard deviation are computed from the recorded $d_{ij}$ values of all detected pairs across all $8{,}000$ Monte Carlo seeds and $200$ snapshot intervals. The agreement between theoretical and empirical values confirms that the uniform-cell assumption is a valid engineering approximation for the Rush-In geometry. Objects move toward targets in the interior of cells, producing slightly higher densities on the inward side, which lowers pair separations relative to the uniform-cell value and accounts for the small undershoot in both empirical parameters.

\begin{table}[ht]
\centering
\caption{Analytically derived and empirically measured parameters of the
pair-distance distribution for the Rush-In scenario at $L = 50$ km.
Empirical values are computed from all detected pairs across $8{,}000$
Monte Carlo seeds.}
\label{tab:gaussian_params}
\begin{tabular}{lcccc}
\hline
\textbf{Parameter} & \textbf{Theoretical} & \textbf{Empirical (DC)} & \textbf{Deviation} & \textbf{Derivation} \\
\hline
$\mu = \bar{d}$ & $33.09$ km & $32.42$ km & $-2.0$\% & Equation~(\ref{eq:mu_robbins}) \\
$\sigma$        & $12.47$ km & $12.06$ km & $-3.3$\% & Equation~(\ref{eq:sigma_derivation}) \\
\hline
\end{tabular}
\end{table}

\subsection{Preservation of \texorpdfstring{$\mathcal{O}(N)$}{O(N)} Complexity}
\label{subsec:complexity}
 
The DC method retains the linear scaling of the original cube algorithm. Computing $\mathbf{I}_i$ and $\mathbf{I}'_i$ for each object requires only scalar division and a floor operation, each $\mathcal{O}(1)$ per object, so assigning all $N$ objects to their respective bins scales as $\mathcal{O}(N)$. For a fixed cell size $L$, the mean number of objects per cell is small and independent of $N$, so the total number of evaluated pairs grows linearly with the population size. Adding a second grid introduces a constant factor of two in the number of bin assignments, which does not alter the asymptotic complexity class. Detection, evaluation, and correction therefore all remain $\mathcal{O}(N)$.

\subsection{Calibration Metrics}
\label{subsec:calibration_metrics}
 
The calibration of each method is assessed using a pairwise reliability diagram. All detected pairs across all Monte Carlo seeds and snapshot intervals are grouped into logarithmically spaced bins by their predicted probability $P_{ij}$. For each bin $b$ containing $n_b$ pairs, the mean predicted probability is
\begin{equation}
\bar{P}_b = \frac{1}{n_b} \sum_{(i,j)\in b} P_{ij},
\label{eq:pave}
\end{equation}
and the empirical collision frequency is
\begin{equation}
f_b = \frac{1}{n_b} \sum_{(i,j)\in b} A_{ij},
\label{eq:fb}
\end{equation}
where $A_{ij} \in \{0,1\}$ is the ground-truth binary collision flag for pair $(i,j)$ at the corresponding snapshot interval. A perfectly calibrated method satisfies $f_b = \bar{P}_b$ for all bins, corresponding to the equality line $f = \bar{P}$ on the reliability diagram (Figure~\ref{fig:reliability}). The Expected Calibration Error (ECE) aggregates the bin-level deviation into a single scalar:
\begin{equation}
\mathrm{ECE} = \sum_b \frac{n_b}{N_{\mathrm{pairs}}} \left|f_b - \bar{P}_b\right|,
\label{eq:ece}
\end{equation}
where $N_{\mathrm{pairs}} = \sum_b n_b$ is the total number of evaluated pairs across all bins.

Because predicted probabilities span approximately two orders of magnitude and calibration errors in this setting are multiplicative rather than additive, the calibration slope is estimated by log-log linear regression

\begin{equation}
  \log_{10}(\bar{P}_b) = m \cdot \log_{10}(f_b) + \log_{10}(A),
  \label{eq:log_regression}
\end{equation}

where the slope $m$ and intercept $A$ are the two regression outputs, with $A$ representing the multiplicative offset in linear scale. The slope $m$ characterizes the probability-dependence of the bias: $m = 1$ indicates that the ratio $\bar{P}_b / f_b$ is constant across all probability bins, while $m > 1$ indicates that the overestimation grows with predicted probability magnitude. The intercept $A$ measures the level of that bias, so that joint calibration requires both $m = 1$ and $A = 1$.\footnote{The intercept is the fitted line evaluated at $f_b = 1$, which lies several decades above the data centroid, so its value is sensitive to the slope estimate and is interpreted here only in conjunction with $m$.} The coefficient of determination $R^{2}$ of the log-log fit is reported as a check on the regression model rather than as a performance measure, since $m$ and $A$ carry no meaning unless the logarithmic relationship is linear. Because pairs assigned $P_{ij} = 0$ cannot enter the log-log regression ($\log_{10}(0)$ is undefined), the slope and intercept reflect only non-blind detections; for the cube method, the regression characterizes co-cell pairs exclusively, and boundary-crossing blind pairs are absent from the fit entirely.

The boundary blindness rate $\beta$ quantifies the fraction of ground-truth collision events for which the algorithm assigns exactly zero predicted risk:

\begin{equation}
\beta = \frac{\sum_{i,j} \!\left[A_{ij} = 1 \;\wedge\; P_{ij} = 0\right]}
             {\sum_{i,j} \!\left[A_{ij} = 1\right]}.
\label{eq:blindness}
\end{equation}

A probability computed at snapshot $t_k$ describes an interval, and that choice affects every metric above. Forward alignment scores $P_{ij}(t_k)$ against events in $(t_k,\, t_{k+1}]$, the interval the probability describes, whereas backward alignment scores it against $(t_{k-1},\, t_k]$, a window the prediction never described. Forward alignment also reads the encounter geometry before the event rather than from a snapshot taken after it has resolved; it matches the no-time-counter convention from which Equation~(\ref{eq:cube_prob}) derives, and it is the convention used by MOCAT-MC. It is adopted throughout this paper.
 
\section{SIMULATION ENVIRONMENT}
\label{sec:experiment}
 
\subsection{The Rush-In Scenario}
\label{subsec:scenario}
 
\begin{figure}[ht]
    \centering
    \subfloat[Initial configuration: objects distributed uniformly
        outside the ROI with inward-directed velocity vectors (gray
        arrows).]{%
        \includegraphics[width=0.32\textwidth]{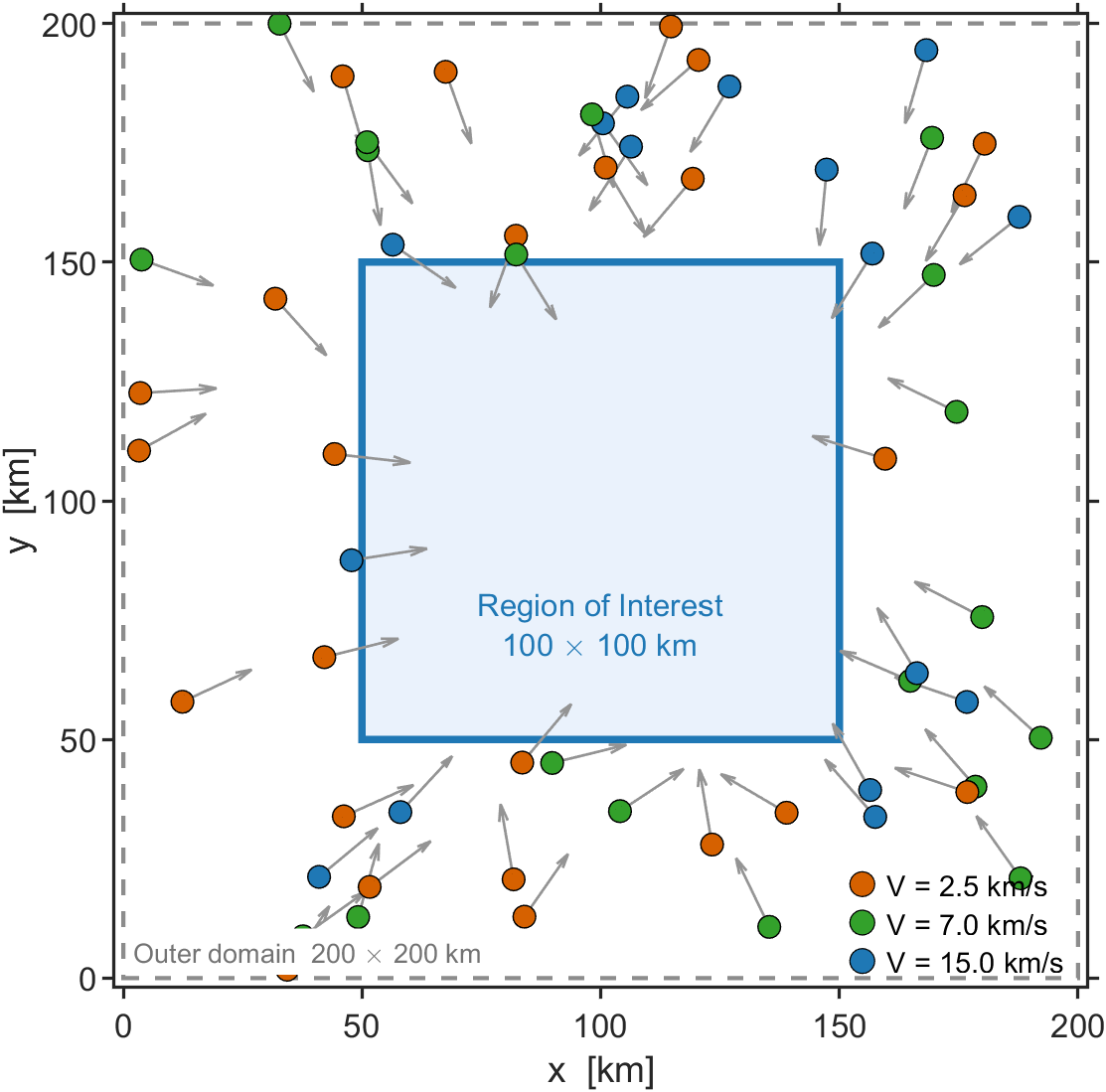}%
        \label{fig:rushin_a}%
    }%
    \hfill%
    \subfloat[$50\times50$~km primary cube grid 
        overlaid.]{%
        \includegraphics[width=0.32\textwidth]{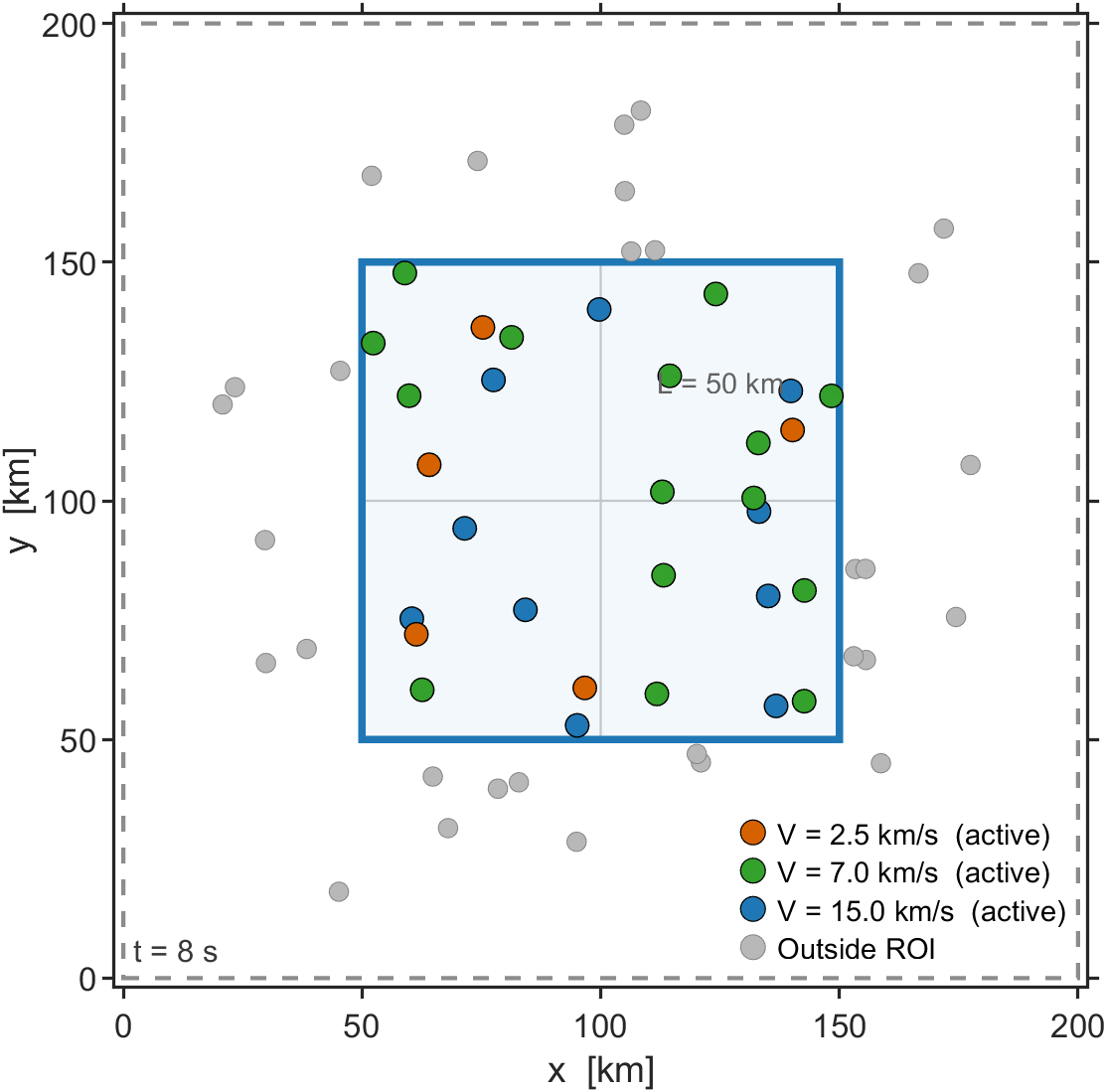}%
        \label{fig:rushin_b}%
    }%
    \hfill%
    \subfloat[Primary cube grid (solid)
    and DC shifted grid (dashed) overlaid inside the ROI.]{%
    \includegraphics[width=0.32\textwidth]{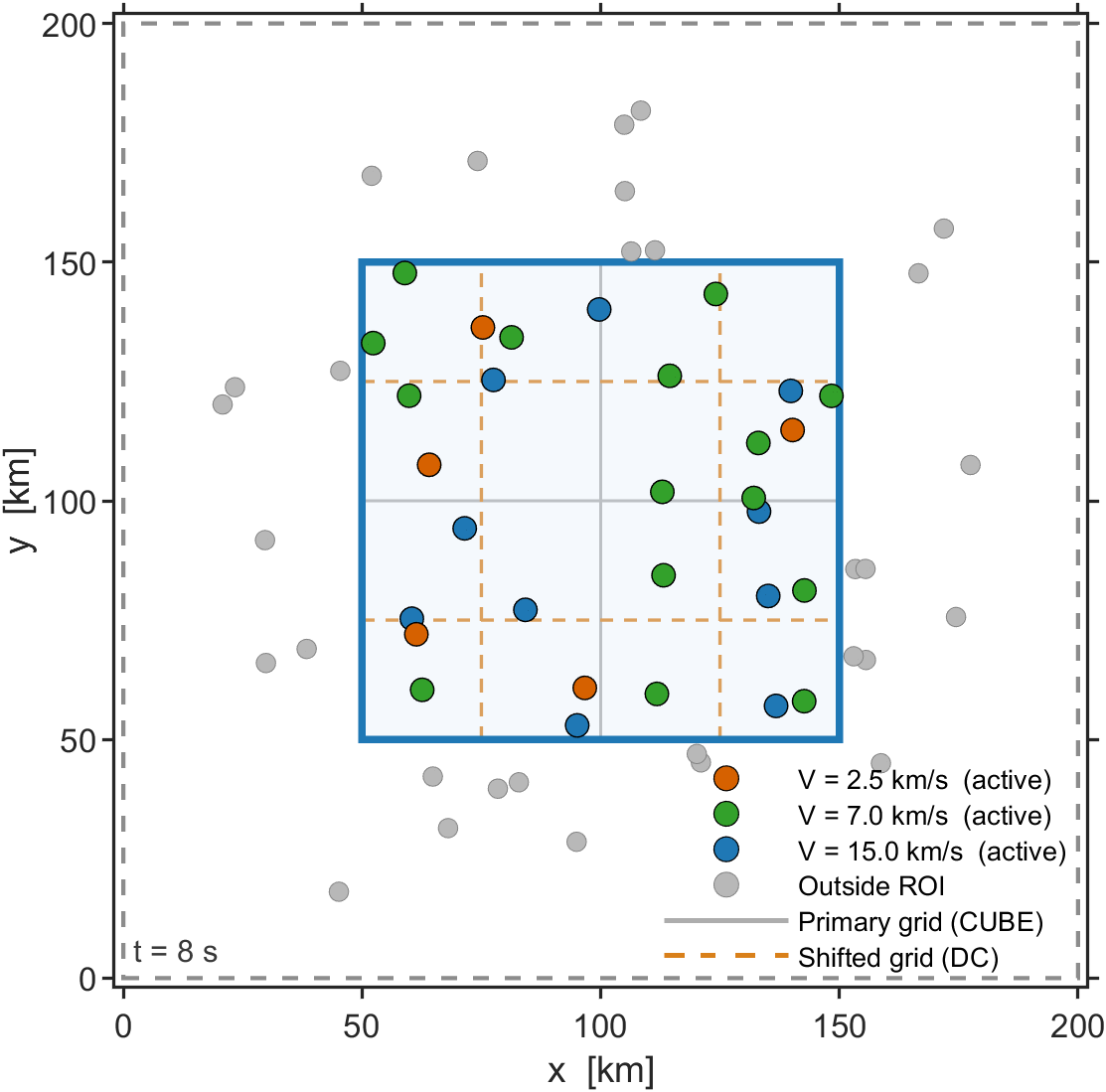}%
    \label{fig:rushin_c}%
}%
    \caption{Rush-In scenario geometry ($N_{\mathrm{obj}} = 200$, $L = 50$~km) and convergence snapshot at $t = 8$~s}
    \label{fig:rushin_scenario}
\end{figure}

The Rush-In scenario is a synthetic convergence benchmark designed to evaluate collision detection performance under controlled geometric conditions. A population of 200 point-mass objects is initialized with positions drawn uniformly at random from a $200 \times 200 \times 200$ km outer domain, with the constraint that each initial position lies outside the $100 \times 100 \times 100$ km region of interest (ROI) spanning $[50, 150]$ km in each Cartesian dimension. Each object is assigned a velocity directed toward a randomly chosen target point inside the ROI, so that all objects converge through the central volume within the simulation window. Object speeds are drawn independently and uniformly from the set $\{2.5,\, 7.0,\, 15.0\}$ km/s and object radii from $\{0.05,\, 0.10,\, 0.20\}$ km, making the population heterogeneous in both size and relative speed. A $5\%$ Gaussian noise term is added to each speed magnitude to break exact synchronization of arrivals. Figure~\ref{fig:rushin_scenario} illustrates the geometry for a representative subset of the population.

The scenario satisfies the two requirements of a calibration benchmark. First, the isotropic uniform initialization is consistent with the kinetic gas assumption underlying Equation~(\ref{eq:cube_prob}), so the Robbins mean prediction of Equation~(\ref{eq:robbins}) can be verified directly against the measured mean pair separation without model mismatch. Second, the controlled-convergence geometry ensures that the encounter event set is dense and reproducible across seeds, yielding both co-cell and boundary-crossing pair detections throughout the simulation window.
 
\subsection{Physics Engine and Ground Truth}
\label{subsec:physics}
Each snapshot interval of $\Delta t_{\mathrm{Cube}} = 0.5$ s is subdivided into 50 physics sub-steps of $\Delta t_{\mathrm{sim}} = 0.01$ s each. The grid snapshot is taken at the start of the interval, and the cube and DC assign predicted probabilities to all detected pairs using the positions at that epoch. The 50 sub-steps then follow: objects are propagated deterministically, and any pair $(i,j)$ whose center-to-center distance falls at or below the combined hard-body radius $R_i + R_j$ is recorded as a physical collision via an elastic momentum-conserving model. The binary ground-truth label
\begin{equation}
A_{ij} =
\begin{cases}
1 & \text{if pair $(i,j)$ collides during the sub-steps following the snapshot,}\\
0 & \text{otherwise,}
\end{cases}
\label{eq:ground_truth}
\end{equation}
is recorded for every pair evaluated at the snapshot, provided both objects lie within the ROI. Because the physics engine operates at the sub-step level and is independent of the grid partitioning, $A_{ij}$ provides an unbiased reference for all calibration metrics defined in the calibration metrics section.
At the Rush-In velocity scale, the mean cell traversal time $\Delta t_{\mathrm{trav}} = L / V_{\mathrm{rel}}$ ranges from 3 to 20 s, so the condition $\Delta t_{\mathrm{Cube}} \ll \Delta t_{\mathrm{trav}}$ holds throughout and objects move little relative to the cell between consecutive snapshots.
 
\subsection{Monte Carlo Protocol}
\label{subsec:mc}

Each seed draws a fresh set of initial object positions from the uniform outer-domain distribution and runs the full 200-snapshot simulation. All detected pairs across snapshots and seeds are accumulated into the reliability diagram bins described in the calibration metrics section.
Two experiments are conducted. The main experiment uses 8,000 seeds for DC and 16,000 for the cube baseline and reports all blindness rates, calibration slopes, and correction results. A second experiment sets $\Delta t_{\mathrm{Cube}} = \Delta t_{\mathrm{sim}} = 0.01$ s so that every physical collision coincides with a snapshot epoch, which removes the temporal source of blindness and isolates the residual to geometric gaps in the detection architecture.
Table~\ref{tab:sim_params} summarizes all simulation parameters.

\begin{table}[ht]
\centering
\caption{Rush-In simulation parameters.}
\label{tab:sim_params}
\begin{tabular}{lll}
\hline
\textbf{Parameter} & \textbf{Symbol} & \textbf{Value} \\
\hline
Cell side length              & $L$                          & 50 km \\
Region of interest            & --                           & $100 \times 100 \times 100$ km \\
Number of objects per seed    & $N_{\mathrm{obj}}$           & 200 \\
Object speed options          & $v$                          & 2.5, 7.0, 15.0 km/s \\
Object radius options         & $R_i,\,R_j$                          & 0.05, 0.10, 0.20 km \\
Total simulation time         & $T_{\max}$                   & 100 s \\
Physics sub-step              & $\Delta t_{\mathrm{sim}}$    & 0.01 s \\
Snapshot interval             & $\Delta t_{\mathrm{Cube}}$   & 0.5 s \\
Snapshots per seed            & --                           & 200 \\
Seeds (Cube baseline)         & --                           & 16,000     \\
Seeds (DC main experiment)    & --                           & 8,000      \\
Seeds (sync.\ experiment)     & --                           & 4,000 \\
\hline
\end{tabular}
\end{table}

\section{RESULTS}
\label{sec:results}
 
\subsection{Boundary Blindness}
\label{subsec:results_blindness}

At $L = 50$ km, the cube assigns zero probability to $7.09\%$ of ground-truth collisions, compared with $0.55\%$ for DC. Setting $\Delta t_{\mathrm{Cube}} = \Delta t_{\mathrm{sim}} = 0.01$ s ensures that every collision occurs at a snapshot epoch, so any event the grids then miss is missed for geometric reasons alone. Under that setting, the cube misses $0.80\%$ of events and DC fewer than $0.01\%$. This decomposes the cube's operational rate into $6.29$ percentage points of temporal blindness and $0.80$ points of spatial blindness, and shows that DC's residual is almost entirely temporal. Table~\ref{tab:blindness} summarizes the rates for both experiments.

\begin{table}[ht]
    \centering
        \caption{Boundary blindness rates under forward alignment at $L = 50$~km
        ($\Delta t_{\mathrm{sim}} = 0.01$~s). The synchronized experiment sets
        $\Delta t_{\mathrm{cube}} = \Delta t_{\mathrm{sim}} = 0.01$~s to isolate
        geometric from temporal blindness.}
    \label{tab:blindness}
    \begin{tabular}{lcc}
        \hline
        Method & $\Delta t_{\mathrm{cube}}$ [s] & $\beta$ \\
        \hline
        Cube                  & 0.50 & 7.09\% \\
        DC                    & 0.50 & 0.55\% \\
        Cube (synchronized)   & 0.01 & 0.80\% \\
        DC (synchronized)     & 0.01 & $<0.01\%$ \\
        \hline
    \end{tabular}
\end{table}
 
\subsection{Mean Collision Separation Validation}
\label{subsec:results_mcs}

The Robbins prediction of Equation~(\ref{eq:robbins}) is validated by computing the empirical mean of $d_{ij}$ across all DC-detected pairs from all 8,000 seeds and 200 snapshot intervals. The measured mean is $32.42$ km against the predicted $\bar{d} = 0.6617 \times 50 = 33.09$ km, a deviation of $-2.0\%$, and the measured standard deviation is $12.06$ km against the predicted $12.47$ km, a deviation of $-3.3\%$. Objects move toward targets in the cell interior, producing slightly higher densities on the inward side, which lowers pair separations relative to the uniform-cell value. The agreement confirms that $\bar{d} = 0.6617\,L$ and $\sigma = 0.2494\,L$ are valid reference parameters for this geometry.
 
\subsection{Calibration and Correction Results}
\label{subsec:results_correction}

\begin{figure}[ht]
\centering
\includegraphics[width=0.8\textwidth]{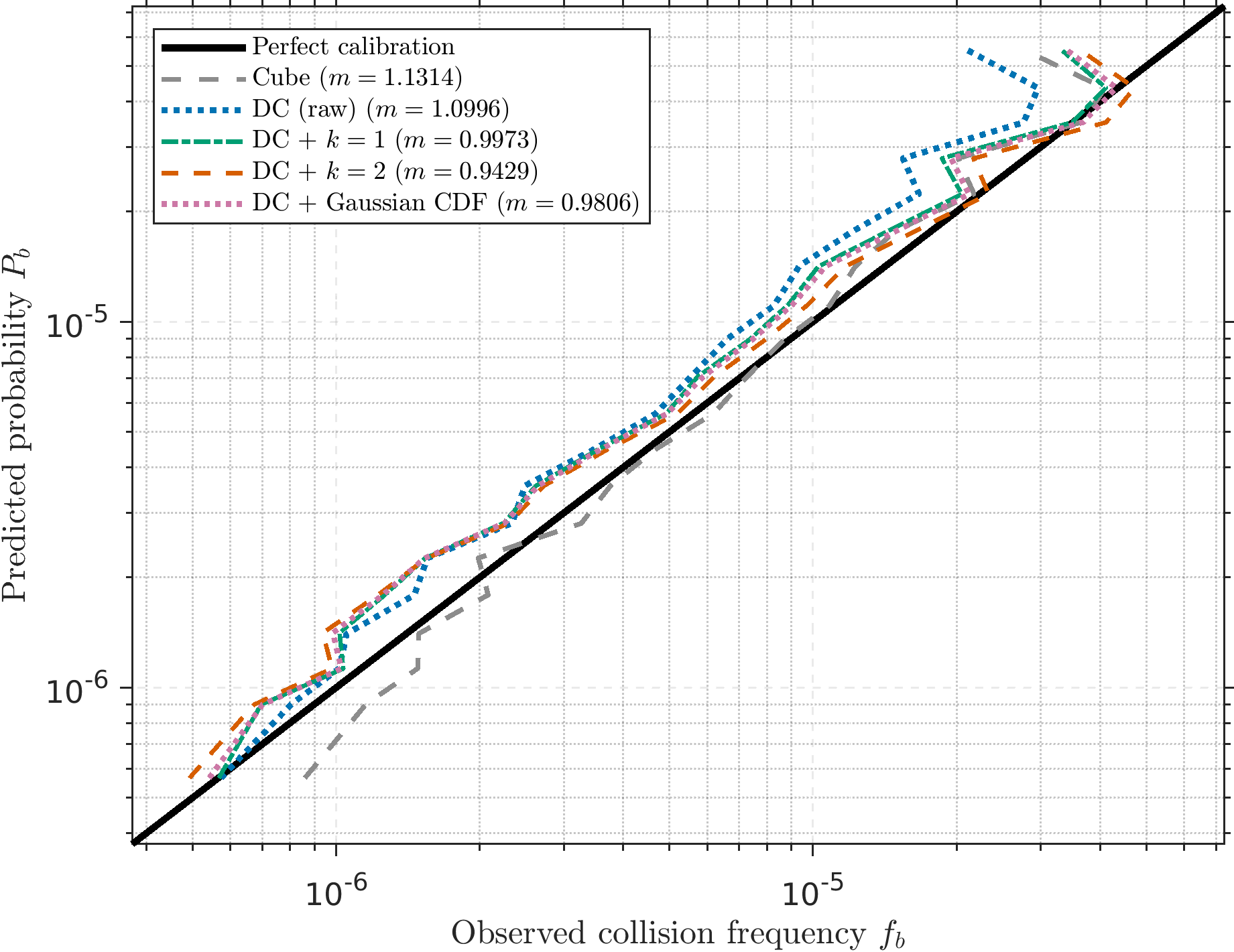}
\caption{Pairwise reliability diagram for all five evaluated methods.}
\label{fig:reliability}
\end{figure}

Figure~\ref{fig:reliability} shows the pairwise reliability diagram for all five evaluated methods, and Table~\ref{tab:calibration} reports the corresponding regression statistics from Equation~(\ref{eq:log_regression}) and calibration metrics. A slope $m > 1$ indicates that the overestimation grows with predicted probability magnitude; $m < 1$ indicates that the correction has been applied too aggressively, driving predictions below the calibration target; and joint $m = 1$, $A = 1$ indicates full calibration in both proportionality and absolute level.

\begin{table}[h]
    \centering
    \caption{Log-log reliability diagram regression statistics from Equation~(\ref{eq:log_regression}) and calibration metrics for each method under forward alignment at $L = 50$~km.}
    \label{tab:calibration}
    \begin{tabular}{lccccc}
        \hline
        Method
            & $m$
            & $A$
            & $R^{2}$
            & $|1-m|$ [\%]
            & ECE [$\times10^{-7}$] \\
        \hline
        Cube                & 1.1314 & 4.839 & 0.9847 & $13.1\%$ over & 6.21  \\
        DC (raw)            & 1.0996 & 4.633 & 0.9864 & $10.0\%$ over & 16.45 \\
        DC $+$ $k=1$        & \textbf{0.9973} & 1.211 & 0.9901 & $\phantom{0}\textbf{0.3\%}$ \textbf{under}  & 10.66  \\
        DC $+$ $k=2$        & 0.9429 & 0.595 & \textbf{0.9903} & $\phantom{0}5.7\%$ under  & 7.73  \\
        DC $+$ Gaussian CDF & 0.9806 & \textbf{0.975} & 0.9901 & $\phantom{0}1.9\%$ under & 9.72  \\
        \hline
    \end{tabular}
\end{table}

Both uncorrected methods are substantially miscalibrated: the cube at $m = 1.1314$ and $A = 4.839$ and the uncorrected DC at $m = 1.0996$ and $A = 4.633$. Neither intercept can be read as a bias factor at these slopes, so no ordering is claimed between the two. The pair demonstrates that recovering boundary-crossing pairs does not by itself improve calibration, since Equation~(\ref{eq:dc_detection}) changes which pairs are evaluated and leaves Equation~(\ref{eq:cube_prob}) untouched. The power-law correction at $k = 1$ gives the closest slope at $0.9973$ but retains an intercept of $1.211$, while $k = 2$ carries both further at $0.9429$ and $0.595$, consistent with the quadratic bound on the no-time-counter bias established by Alexander et al.~[\citen{alexander1998,alexander2000erratum}]. The Gaussian CDF correction of Equation~(\ref{eq:cdf_correction}), derived from the pair-distance geometry without reference to kinetic theory or to any quantity estimated from the data, falls between the two exponents on both metrics at $m = 0.9806$ and $A = 0.975$.

Judged jointly on $m$ and $A$, the Gaussian CDF correction is the only configuration within $3\%$ of unity on both, and no other configuration is within $20\%$ on both. The power-law correction at $k = 1$ achieves the better slope in isolation but retains a $21\%$ intercept error, which is why the two metrics must be read together rather than ranked separately.

Equation~(\ref{eq:ece}) weights bins by occupancy and measures absolute deviation, which rewards suppression of the low-probability bins that hold most of the pairs. The cube therefore records the lowest value while assigning zero probability to $7.09\%$ of true events, and DC raw records the highest while evaluating the most pairs. ECE is reported as a complementary measure of average absolute deviation in linear space, and the slope and intercept remain the primary calibration metrics.

\section{APPLICATION TO ORBITAL CAPACITY}
The pairwise reliability diagram is validated against a deterministic physics engine in the controlled Rush-In geometry, characterizing per-pair calibration under isotropic conditions, but does not establish whether the corrections recover agreement with the absolute collision rate observed under realistic orbital propagation. Facchinetti~[\citen{facchinetti2024cube}] compared the standard cube method against a deterministic propagation benchmark and found that it does not reproduce the reference collision rate. The journal version of this work addresses whether the corrections derived here recover agreement with an absolute deterministic benchmark in the orbital debris environment.

\begin{figure}[ht]
    \centering
    \subfloat[Total object population over 50 years; shaded bands
        show $\pm 1\sigma$ across 4{,}000 seeds.]{%
        \includegraphics[width=0.8\textwidth]{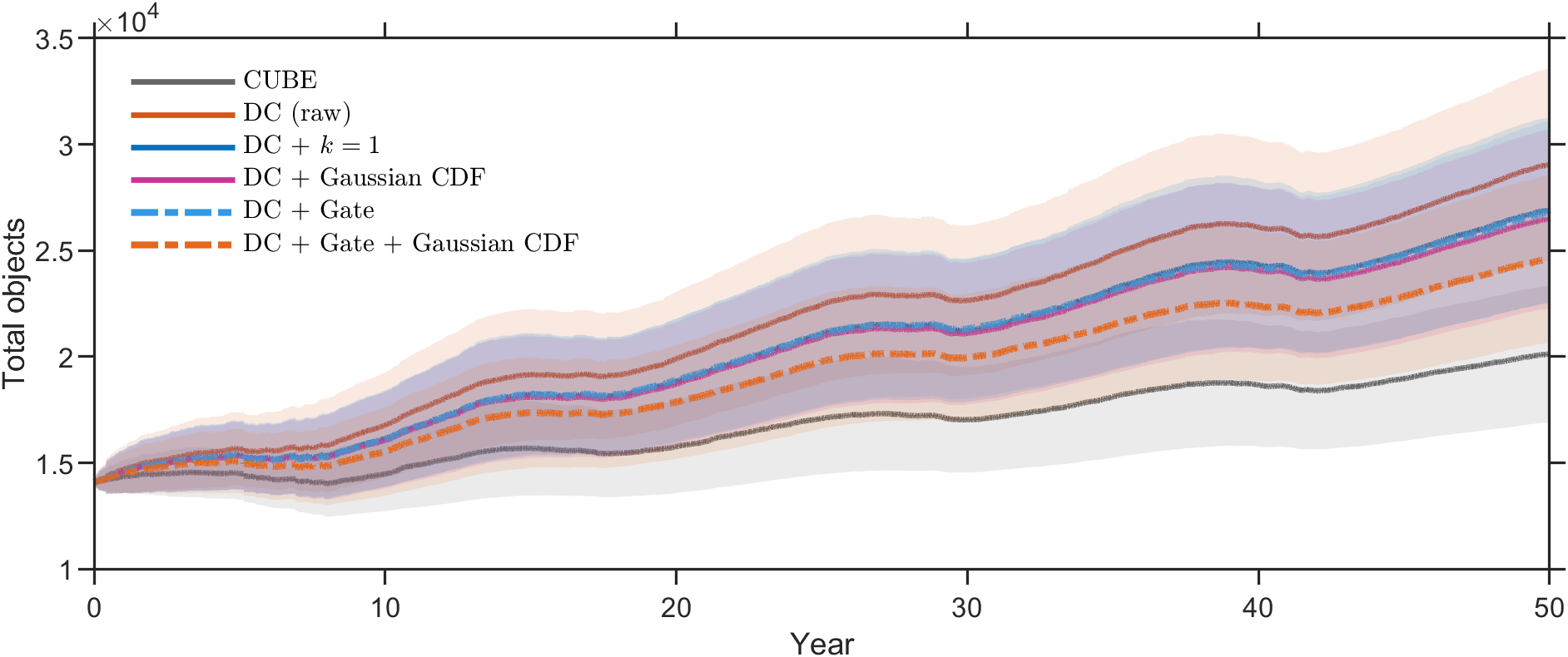}%
        \label{fig:mocat_pop_a}%
    }\\[2pt]
    \subfloat[Debris fragment count over 50 years.]{%
        \includegraphics[width=0.8\textwidth]{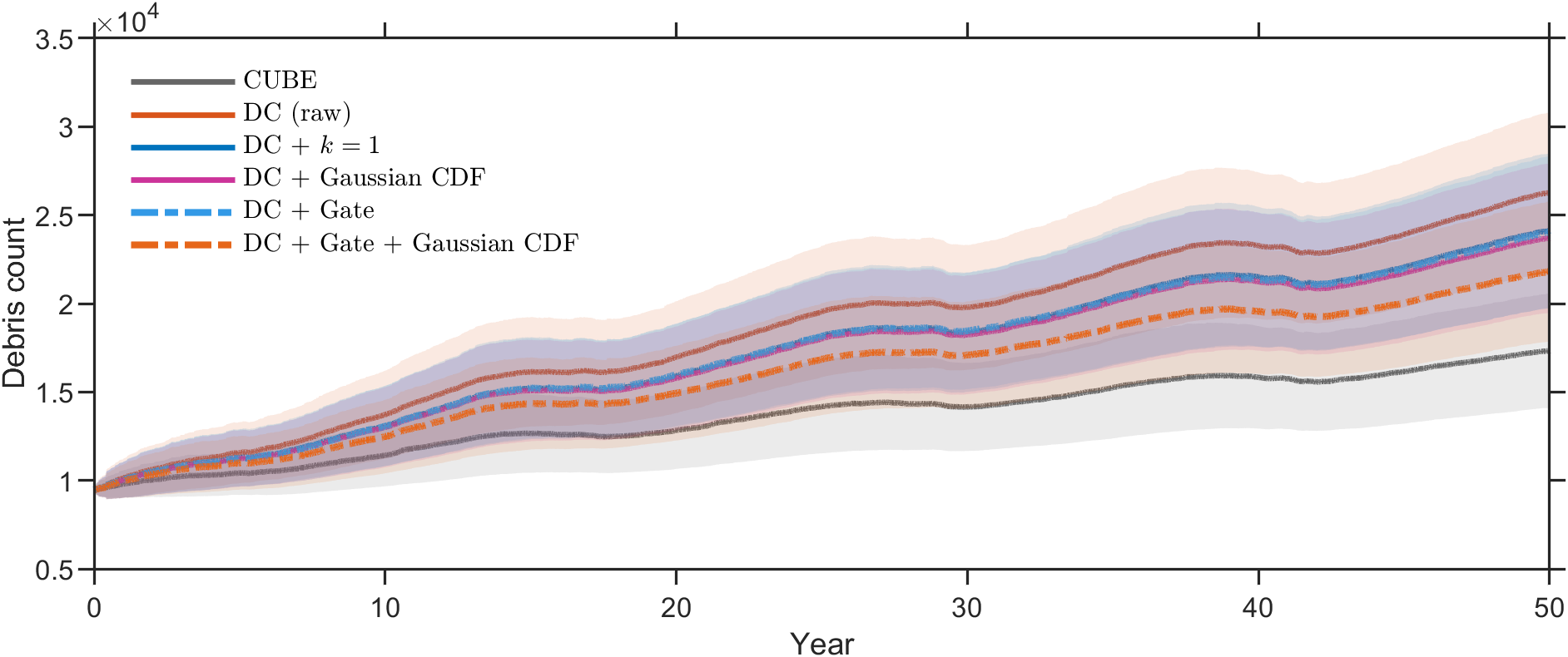}%
        \label{fig:mocat_pop_b}%
    }%
    \caption{Representative 50-year MOCAT-MC projections for six correction configurations (4{,}000 seeds each, $L = 50$~km).}
    \label{fig:mocat_population}
\end{figure}

In the MOCAT-MC 50-year ensemble, the DC method evaluates $541{,}006$ conjunctions per seed against $180{,}237$ for the cube, and the resulting collision rate rises from $23.88$ to $55.14$ events per seed. The excess arises because the $L/2$-shifted secondary grid places objects from adjacent altitude shells in the same shifted cell even when their orbital altitude ranges are fully disjoint. Kessler~[\citen{kessler1978collision,kessler1981derivation}] showed that the spatial density of an orbiting object is zero outside the interval $[r_p,\, r_a]$, where $r_p = a(1-e)$ and $r_a = a(1+e)$ are the perigee and apogee radii; it follows from his integral collision rate that two objects whose altitude bands do not overlap have exactly zero collision rate regardless of their instantaneous separation $d_{ij}$. A radial range overlap gate enforces this condition by computing the band gap $d_{\mathrm{gap}} = \max\!\bigl(0,\, \max(r_{p,i}, r_{p,j}) - \min(r_{a,i}, r_{a,j})\bigr)$ and setting $\alpha_{\mathrm{RO}} = 1$ when $d_{\mathrm{gap}} \le R_i + R_j$ and $\alpha_{\mathrm{RO}} = 0$ otherwise, using only the semi-major axis $a$ and eccentricity $e$ already carried in the MOCAT-MC state vector at $\mathcal{O}(1)$ cost per pair. This gate has been implemented in two additional configurations: one applies it to DC raw detections without any per-pair correction, and the other combines it with the Gaussian CDF correction in Equation~(\ref{eq:cdf_correction}).

The DC method, both per-pair corrections and the radial-range overlap gate, has been implemented in MOCAT-MC and applied to 50-year ensemble simulations at 4{,}000 Monte Carlo seeds per configuration. Figure~\ref{fig:mocat_population} shows representative projections for the cube baseline, DC raw, DC $+$ Gate, DC $+ k = 1$, DC $+$ Gaussian CDF, and DC $+$ Gate $+$ Gaussian CDF. The projected population falls in that order, so the gate alone suppresses less than either per-pair correction, and the two act together in the gated Gaussian configuration, which lands the lowest of the five DC variants.

\section{CONCLUSION}
\label{sec:conclusion}

This paper introduced the Double Cube (DC) method, which recovers boundary-crossing conjunctions through a spatially shifted secondary grid using bin-index lookup alone, preserving $\mathcal{O}(N)$ complexity. Validated against a deterministic physics engine in an isotropic Rush-In convergence scenario at $L = 50$ km, DC reduces the blindness rate from $\beta_{\mathrm{Cube}} = 7.09\%$ to $\beta_{\mathrm{DC}} = 0.55\%$, and a synchronized experiment with $\Delta t_{\mathrm{Cube}} = \Delta t_{\mathrm{sim}}$ leaves the cube at $0.80\%$ and DC below $0.01\%$, showing that DC's residual is almost entirely temporal in origin.

Recovering those pairs does not improve calibration on its own, because the detection rule and the probability formula fail independently. Both uncorrected methods remain miscalibrated, at $m = 1.1314$ for the cube and $m = 1.0996$ for DC. The power-law correction moves the slope to $0.9973$ at $k = 1$ and $0.9429$ at $k = 2$, and the Gaussian CDF correction, derived entirely from the mean $\bar{d} = 0.6617L$ and standard deviation $\sigma = 0.2494L$ of the pair-distance distribution under the uniform cell assumption, reaches $m = 0.9806$ with an intercept of $0.975$ and is the only configuration within $3\%$ of unity on both. Neither moment is fitted to simulation data.

Representative 50-year ensemble projections and the radial range overlap gate implementation are presented in the preceding section. The closed-form detection volume of the dual grid, the population-level comparison against a deterministic twenty-year cascade propagation of the low Earth orbit catalog, and the limits of calibration evidence generated inside a Bernoulli-sampled model are treated in the journal version of this work.

\bibliographystyle{AAS_publication}  
\bibliography{references}   

\end{document}